%% file: main.tex
\documentclass[preprint]{vgtc} 

\ifpdf
  \pdfoutput=1\relax                   
  \usepackage{graphicx}                
  \DeclareGraphicsExtensions{.pdf,.png,.jpg,.jpeg} 
\else
  \usepackage{graphicx}                
  \DeclareGraphicsExtensions{.eps}     
\fi%

\graphicspath{{figures/}{pictures/}{images/}{./}} 

\usepackage{microtype}                 
\PassOptionsToPackage{warn}{textcomp}  
\usepackage{textcomp}                  
\usepackage{mathptmx}                  
\usepackage{times}                     
\usepackage{cite}                      
\usepackage{tabu}                      
\usepackage{booktabs}                  
\usepackage{float}

\usepackage{adjustbox}
\usepackage{xr}

\vgtccategory{Research}

\ieeedoi{10.1109/TVCG.2023.3320242}

\renewcommand \manuscriptnotetxt 

\title{Exploring Users Pointing Performance on Large Displays with Different Curvatures in Virtual Reality}

\author{\authororcid{A K M Amanat Ullah}{0000-0001-5402-0160} \thanks{e-mail: amanat7@student.ubc.ca}\\ %
        \scriptsize University of British Columbia \\\scriptsize Okanagan, BC, Canada %
\and \authororcid{William Delamare}{0000-0002-1830-4294} \thanks{e-mail: william.delamare@acm.org}\\ %
     \scriptsize Univ. Bordeaux, ESTIA-Institute of Technology, EstiaR \\\scriptsize F-64210 Bidart, France%
\and \authororcid{Khalad Hasan}{0000-0002-4815-5461} \thanks{e-mail: khalad.hasan@ubc.ca}\\ %
     \scriptsize University of British Columbia \\\scriptsize Okanagan, BC, Canada}

\newcommand{\degree}{\textbf{$^{\circ}$}}
\newcommand{\update}[1]{{\textcolor{black}{#1}}} 

\teaser{
  \centering
  \includegraphics[width=\linewidth]{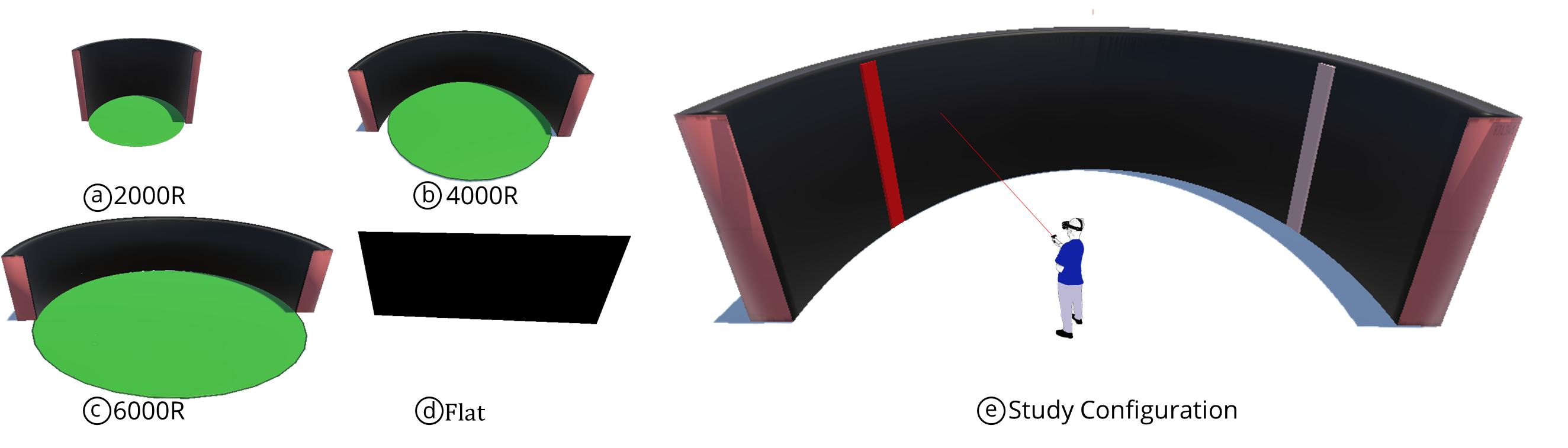}
  \caption{Display Curvature: (a) 2000R, (b) 4000R, (c) 6000R where the numbers refer to screen radius in mm. (d) A flat display. (e) A user is using raycast for pointing on a 4000R display}
  \label{all5displayTypes}
  \label{fig:teaser}
}

\abstract{
Large curved displays inside Virtual Reality environments are becoming popular for visualizing high-resolution content during analytical tasks, gaming or entertainment. 
Prior research showed that such displays provide a wide field of view and offer users a high level of immersion. 
However, little is known about users’ performance (e.g., pointing speed and accuracy) on them.
We explore users' pointing performance on large virtual curved displays.
We investigate standard pointing factors (e.g., target width and amplitude) in combination with relevant curve-related factors, namely display curvature and \update{both linear} and angular measures. 
\update{Our results show that the less curved the display, the higher the performance, i.e., faster movement time.}
This result holds for pointing tasks controlled via their visual properties (linear widths and amplitudes) or their motor properties (angular widths and amplitudes). 
\update{Additionally, display curvatures significantly affect the error rate for both linear and angular conditions.}
Furthermore, we observe that curved displays perform better or similar to flat displays based on throughput analysis.
Finally, we discuss our results and provide suggestions regarding pointing tasks on large curved displays in VR. }


\begin{document}
\maketitle

\input{1_introduction}

\input{2_background}


\input{4_curve_Exploration-Designspace}

\input{6_exp-protocol}

\input{7_exp-results}


\input{10_discussion}
\input{11_conclusion}


\acknowledgments{
 This research was partially funded by a Natural Sciences and Engineering Research Council (NSERC) grant. We thank Joel Thiessen and Garth Evans from UBC Okanagan’s VISUALIZATION+ EMERGING MEDIA STUDIO (VEMS) for providing access to the large curved display facility. We appreciate our colleagues: Sohan Chowdhury, Omang Baheti, and Marium-E- Jannat for their support and insights.}

\bibliographystyle{abbrv-doi-hyperref}
\bibliography{main}
\newpage

\end{document}

%% file: 1_introduction.tex
\section{Introduction}

Large physical displays are becoming popular for visualizing high-resolution content for data visualization \cite{andrews2011information}, for collaboration \cite{novak2008designing} and for entertainment purposes \cite{ardito2015interaction}.
Large displays, with varying curvatures (e.g., flat and curved display), offer users detailed information and increase their task performance and satisfaction \cite{czerwinski2003toward, tan2003similar}.
However, such displays require large space and advanced hardware and are hence scarcely used by general audiences \cite{cavallo2019dataspace, Ullah2023vrvsreal}. With the recent advancement of Head Mounted Displays (HMDs), large displays in Virtual Reality (VR) are becoming a potential alternative for visualizing high-resolution data \cite{cao2019large, Ullah2023vrvsreal}. 
Although extensive research has been carried out on large physical displays, very little is known about users' performance (e.g., pointing speed and accuracy) on large virtual displays with various properties such as display curvatures.
Thus, advancements in this field warrant a closer examination of users' performance while interacting with large virtual displays in VR.

Previous work widely explored users' performance on flat large displays via Fitts pointing tasks \cite{nancel2013high, haque2015myopoint, lischke2016magic, kopper2010human}. 
These studies showed that linear target amplitude (i.e., the linear distance between targets) and linear target width significantly affect users' pointing performance on large physical flat displays \cite{tao2021freehand, lischke2016magic, haque2015myopoint, siddhpuria2018pointing, nancel2013high}. 
Previous work also revealed that curved displays offer improved user performance compared to flat displays as they are free from region bias: the entire display is situated uniformly around users, which mimics their natural field of view \cite{kyung2021curved, shupp2009shaping}. 
Although extensive research has been carried out on users' pointing performance on large flat displays, the effect of curved display-related properties (e.g., different display curvature) while pointing has not been explored due to hardware and cost-related constraints \cite{cavallo2019dataspace, Ullah2023vrvsreal}.
We argue that virtual environments offer an opportunity to initially study curved displays without the limitations of physical prototypes.

In this paper, we investigate users' pointing performance on large virtual displays considering standard pointing factors (e.g., target width and movement amplitude) in conjunction with curve-related factors (e.g., display curvatures). 
In our user study, we explore the effect of different display curvatures on users’ pointing performance with 1D Fitts' law tasks.
For a thorough exploration, we also consider (i) angular and linear measurements for experimental conditions as not to favor any display type and (ii) different Fitts model variations each associated to different distal pointing tasks in VR theoretical representations (e.g., standard or peephole pointing). 
\update{Results reveal that less curved displays have higher performance when considering movement time for both the linear and the angular measurement tasks.}
The results hold for both angular and linear conditions and for all Fitts model variations - with each model displaying a good fit to model distal pointing task in VR (i.e., most $R^2$ > 0.9).
\update{Furthermore, less curved displays have lower error rates for the angular condition whereas more curved displays have lower error rates for the linear condition.} 
\update{From our user study, we found that more curved displays required the participants to perform increased body movement for the linear condition - leading to slower selections.}
\update{For the linear condition, the width of the target changed based on the display curvature; the less curved the display is, the target is projected farther away.
This leads to higher visual target sizes for the less curved displays, allowing faster selection and higher accuracy than more curved displays.}
While considering Fitts law's throughput criteria,  we observe that curved displays have similar or even better performance than Flat displays.


The contributions of the paper are:
\begin{itemize}
\item A comprehensive exploration of the effect of display curvatures on pointing performance on large virtual displays in VR with:
    \begin{itemize}
        \item the comparison of linear and angular measurements to define experimental conditions
        \item the validation of three Fitts law variations (standard/one-part, two-part, and peephole) for distal pointing in VR
    \end{itemize}
\end{itemize}

%% file: 2_background.tex
\section{Background and Related Work}
We first consider prior work investigating users' pointing performance on large flat and curved displays.
We then review works focusing on pointing tasks in head-mounted displays.

\subsection{Flat and Curved Large Displays}

Researchers have explored users pointing performance on large flat displays. 
There has been little research focusing on pointing performance on curved displays, and no research focusing on pointing on virtual curved and flat screens.

\subsubsection{Pointing Techniques on Large Flat Displays}
Ray-casting (i.e., a ray is projected from the tip of an input device) is the most widely used method for pointing at targets on large displays \cite{cavens2002interacting, davis2002lumipoint, oh2002laser, jota2010comparisonGIpaper}. 
Jota et al. \cite{ jota2010comparisonGIpaper} compared a set of ray-casting techniques and showed that regular laser pointing (projection from the tip of the index finger) leads to higher throughputs and lower fatigue for 1D pointing (i.e. either horizontal or vertical movements) than the other explored techniques.
Haque et al. \cite{haque2015myopoint} explored \update{Myopoint} which leverages a Myo armband to track hand movements to point at 2D objects displayed on a large flat display.
They showed that \update{Myopoint} allows the selection of 48 mm targets with a 15\% error rate. 
Nancel et al. \cite{nancel2013high} proposed a technique Conthead that combines head orientation for coarse pointing with touch-based relative pointing. 
They showed that the technique could acquire targets as small as 4.16 mm on a 5.5-meter (resolution 22080$\times$7360) display.
Siddhpuria et al. \cite{siddhpuria2018pointing} compared seven device-based pointing techniques (i.e., using a watch or a smartphone) for distal pointing tasks on a large flat display. 
They found that relative touch pointing with two hands had the best performance on smartphones.
In summary, prior work extensively explored different device-based and hand-based pointing techniques on large flat displays. 
Instead of exploring both devices and hand-based pointing, we only considered devices-based pointing (i.e., with controller) in large virtual displays and focused on exploring users' pointing performance on these displays.

\subsubsection{Large Curved Displays}
We found the following two predominant research streams on large curved displays.

\textbf{Flat versus Curved Large Displays:}
Shupp et al. \cite{shupp2009shaping} compared a large flat display and a curved display for different geospatial tasks.
They showed that curved displays i) lead to higher performances than flat displays, ii) suffered less from region bias effects (i.e. position of elements on display), and iii) that they facilitated both search (i.e., finding an icon in a map) and comparison (i.e., comparing between values from static visual data) tasks. 
However, users had to turn around on themselves more compared to flat displays.
In contrast, flat displays offered a better overview than curved displays and were preferred for tasks involving gaining insights from static visual data. 
However, flat displays forced users to walk more and sustained region bias effects as users focus more on the center of the display than on the edges, unlike curved displays uniformly distributed around users.
Ahn et al. \cite{ahn2014research} investigated differences between a 55" flat and a 55" curved display regarding focused point, posture, satisfaction, and tilt angles. 
They showed that participants preferred to watch content on curved displays and participants' point of focus differed between curved and flat displays.
Overall, prior research showed that curved displays offer a wide range of benefits to flat displays.
However, these works did not examine curved display properties (e.g., display curvature) that influence users' performance. 

\textbf{Curved Displays Specificities:}
Due to their improved ergonomics, curved screens have been proven to be more suitable for visual tasks compared to a flat display \cite{urakami2021comparing, ahn2014research, kyung2021curved, choi201553, shupp2009shaping}. 
Kyung and Park \cite{kyung2021curved} compared 33" and 50" displays with different display curvatures (i.e., 40 cm, 60 cm, 120 cm radius) and recommended using a 60 cm radius for both displays for visual search tasks.

However, there has been limited research focusing on pointing tasks on large curved displays. 

To our best knowledge, the only pointing research on curved displays consisted of a Fitts law study by Hennecke et al. \cite{hennecke2013investigating} and Ens et al. \cite{ens2016moving}.
Hennecke et al. \cite{hennecke2013investigating} considered target width, amplitude, direction and display region for 2D pointing tasks with direct touch and mouse inputs on a 42" curved display. 
Results showed the Fitts law applies to curved displays irrespective of input devices.
Their display, however, was neither fully curved nor large in size due to hardware restrictions as they used two HD projector outputs to create an ergonomically optimized curved display.
Ens et al. \cite{ens2016moving} conducted a 1D Fitts law task in a cave environment using targets arranged in a circular orientation while considering five different field-of-view conditions (i.e., 8\degree, 16\degree, 32\degree, 64\degree, 128\degree, and no restrictions). 
Although they observed a trend of improved performance with higher field-of-view, the effect of field-of-view was not statistically significant for error and movement time.

In summary, pointing performance on large curved displays has received little attention. 
In addition, the study on curved displays did not account for factors unique to curved displays, such as the curvature of the displays.
Furthermore, research on pointing performance has only been conducted for physical displays, not virtual ones, thus warrants further investigation.

\subsection{Target Pointing in HMD}
Researchers explored target pointing tasks with Fitts in virtual environments.
For instance, Ullah et al. \cite{Ullah2023vrvsreal} conducted a study on Fitts' law target pointing, comparing a large curved display in VR with a physical large curved display. Their results indicated that virtual curved displays could serve as alternatives to physical ones since there was no difference in performance (i.e., movement time and error rate) between the two platforms. However, their study was limited to a single display curvature (i.e., 3270R) and did not consider the effect of display curvature on pointing performance.
Shi et al. \cite{shi2022pointing} ran a Fitts' law study to find the effect of input types (i.e., raycast and virtual touch) and other factors (e.g., user distance, target width) on users pointing performance in a head-mounted VR environment. 
Results showed that movement time was significantly affected by input type, user distance, and target width. 
Hansen et al. \cite{hansen2018fitts} explored users' pointing performance for three input modalities (i.e., gaze, head tracking, and mouse) in head-mounted displays where they found that the mouse was the fastest followed by head pointing and gaze pointing. 
They also found that the perceived workload with gaze pointing was lower than with head pointing. 
\update{Batmaz et al. \cite{batmaz2022effect} explored distal pointing and found that targets placed at the focal distance of a VR headset (0.75m in the case of HTC VIVE) have higher throughput compared to target placed outside the focal distance (further than 0.75m).}
Batmaz et al. \cite{batmaz2019head} used Fitts' law for pointing tasks in AR and VR where users had to move a motion-tracked wand to point at virtual objects in two movement directions: view direction (i.e., front-to-back movements in the forward direction of where the user is facing) and lateral direction (i.e., pointing left and right).
They showed that users' movement direction and target distance significantly affect users' pointing performance. 
In addition, movements toward the user were quicker than movements away from the user. 
They did not notice any difference in pointing between AR and VR headsets - which contradicts  \cite{jones2008effects,naceri2010depth}.
All the previous studies evaluated pointing performance \cite{batmaz2022effect, batmaz2019head, hansen2018fitts, chowdhury2022wriarm, shi2022pointing} in a 3D environment rendered via an HMD.
We did not find any work focusing on virtual surfaces (flat or curved) rendering interactive and selectable 2D content such as a virtual display inside the virtual environment.

\subsection{Summary}

Previous work extensively explored pointing performances on large flat displays \cite{kopper2010human, shoemaker2012two, janzen2016modeling} where very little is known about users pointing performance on large curved displays.
In addition, very few studies explored properties related to curved-displays, such as display curvature \cite{shupp2009shaping, kyung2021curved}.
Indeed, varying physical properties of large curved displays involves several practical difficulties (e.g., cost, equipment) \cite{cavallo2019dataspace, Ullah2023vrvsreal}.
The fact that 2D content can be displayed in virtual environments (e.g., for visual analytic tasks \cite{cao2019large, Ullah2023vrvsreal}) opens new opportunities.
Yet, previous work with HMD focused on 3D selection task \cite{shi2022pointing, batmaz2019head, hansen2018fitts}, not with virtual displays - flat or curved.
This leaves a complete field of research unexplored: the exploration of large flat and curved displays offered by VR with HMD.

%% file: 4_curve_Exploration-Designspace.tex
\section{Design Space for Curved Displays Exploration}

We first briefly review several Fitts law variations with their usage context.
We then describe the design factors we intend to explore to evaluate pointing performances on virtual curved large displays via a Fitts law study.
These factors include standard pointing factors (e.g., target size and movement amplitude) as well as curved-display specific factors (e.g., curvature angle).

\subsection{Fitts Law variations}
\update{Fitts law, recommended by ISO 9241-9 \cite{iso20009241}, is widely adopted in the research community to model and quantify pointing performances.}
Fitts law is a predictive model of human motor behavior that is applicable for pointing in 1D \cite{fitts1954information, card1978evaluation, jota2010comparisonGIpaper}, and later extended to 2D \cite{mackenzie1992fitts, hoffmann1995effective, hoffmann2011performance, nancel2013high} and 3D pointing selection tasks \cite{triantafyllidis2021challenges, hoffmann2011performance, kulik2020motor, leusmann2021literature}. 
In addition, previous works proposed several extensions of Fitts law which we discuss below:

\subsubsection{One-part Model} The common form of Fitts law (ISO 9241-9 \cite{iso20009241}) for pointing tasks was proposed by MacKenzie \cite{mackenzie1992fitts} which is also known as the Shannon-Fitts formula. 
This formula allows to predict the pointing movement time MT: 
\begin{equation} \label{eqn_Fittslaw_popular}
MT=a+b \log _{2}\left(\frac{A}{W}+1\right)
\end{equation}
Where A represents the movement amplitude, W the target width, and a and b are empirically evaluated constants.
The Index of Difficulty ID of a target corresponds to the logarithmic term $\log _{2}\left(\frac{A}{W}+1\right)$.

 Previous work suggested several extensions of Fitts law to better model different conditions such as including target height \cite{crossman1956measurement, mackenzie1992extending, hoffmann1994effect}, including movement directions in degrees \cite{kondraske1994angular, kopper2010human, kopper2011understanding}, or considering 3D environments  \cite{grossman2004pointing} to name a few.
Next, we describe two variations, the \emph{Two-parts model} \cite{janzen2016modeling, shoemaker2012two} and the \emph{Peephole model} \cite{rohs2008target, cao2008peephole, jin2020modeling}, which conceptually fit to our HMD virtual pointing use-case.

\subsubsection{Two-part Model}
Kopper et al. \cite{kopper2010human, kopper2011understanding} showed that angular width and angular amplitude significantly affect the difficulty of a distal pointing task (i.e., pointing at a target from a distance).
However, Shoemaker et al. \cite{shoemaker2012two} and Janzen et al. \cite{janzen2016modeling} proposed a \textbf{two-part model Fitts law model} to separately consider the effects of width and amplitude.
Authors showed that the two-part model provided a better fit than using angular measurements.
The two-parts model (also called Shanon-Welford formula \cite{shoemaker2012two, janzen2016modeling}) 
predicts movement time via:
\begin{equation} \label{eqn_Fittslaw_twopart}
MT=a+b_1 \log _2(A+W)-b_2 \log _2(W) \\
=a+b \log _2\left(\frac{A+W}{W^k}\right)
\end{equation}

The index of difficulty depends on the target amplitude (A) and target width (W). 
K is a constant to empirically set based on the pointing environment variables, such as cursor gain and user distance from the display. 
a and b are constants - derived from regression analysis. 
Since the two-part model provides a better fit than using angular measurements, we consider this model in our analysis.

\subsubsection{Peephole Model}

The peephole model allows the prediction of movement times based on a two-stage pointing \cite{rohs2008target, cao2008peephole}: a first stage to get the target on screen (e.g., in the field-of-view of an HMD), and a second stage to point at the visible target.
Originally, Rohs and Oulasvirta \cite{rohs2008target} proposed a magic lens pointing model for peephole interfaces on smartphones. 
They described magic lenses as mobile devices with cameras used in the AR environment, which act as a movable window in the virtual space. 
Jin et al. \cite{jin2020modeling} extended the formulas of Magic Lens \cite{rohs2008target} and Peephole pointing \cite{cao2008peephole} to an AR context on a smartphone. 
Interestingly, the peephole pointing model can be applied to HMD and their limited field-of-view similar to the display size S of handheld devices \cite{ens2016moving}. 
The corresponding Fitts law formula we consider is \cite{cao2008peephole}:
\begin{equation} \label{eqn_Fittslaw_Peephole}
 MT=a+b \left( n \log _2\left(\frac{A}{S}+1\right)+(1-n) \log _2\left(\frac{A}{W}+1\right) \right)
\end{equation}

The proposed extensions add a new factor: the display size S. The Index of difficulty ($ID= n \log _2\left(\frac{A}{S}+1\right)+(1-n) \log _2\left(\frac{A}{W}+1\right) $) depends on display size ($S$), target amplitude ($A$), and target width ($W$). 
$n$ is a constant empirically determined and reflects how much the Fitts ID (task difficulty) depends on the display size.

\subsection{Linear and Angular Width and Movement Amplitude}

It is customary to vary target width W and movement amplitude A to cover a wide range of IDs (usually between 2 bits and 8 bits according to the ISO 9241-9 standard \cite{soukoreff2004towards,iso20009241}).
Previous work mostly considered linear target width and movement amplitude, i.e., measured in meters or pixels \cite{ fitts1992information, tao2021freehand, lischke2016magic, haque2015myopoint, siddhpuria2018pointing, nancel2013high}. 
Angular target width and angular amplitude, i.e., measured in degrees or radians, have been introduced to capture distal pointing characteristics \cite{kondraske1994angular, kopper2010human, kopper2011understanding}. 
The choice between linear and angular is usually revealed after the experiment to compare how models fit to empirical data.
\update{For instance, Kopper et al. \cite{kopper2010human, kopper2011understanding} and Jota et al. \cite{jota2010comparisonGIpaper} used linear measurements in the task design (e.g. experimental amplitude A=1.379m from interaction distance, D= 1m), and then used the angular to linear formula by Kopper et al. \cite{kopper2010human} for the result analysis (angular amplitude $\alpha$=69.17°).}
While the analysis and the model comparison can be performed \textbf{after} the experiment, we still need to decide on a measure unit to fix experimental conditions \textbf{before} running the experiment.
Indeed, IDs ($\log _{2}\left(\frac{A}{W}+1\right)$) won't be impacted by a unit transformation because of the ratio A/W.
But the task itself will be different across conditions when considering visual and motor spaces.
As an example, let's consider a condition with A = 2 m, W = 0.1 m (ID = 4.39), and a distance D of 2 m.
This leads to angular measures of A = 20 $\degree$, and W = 1 $\degree$ (ID = 4.39, same as linear as they have same ratio between A and W).
Yet, varying other factors will completely change the selection difficulty users have to perform:
\begin{itemize}
    \item With linear measures, increasing the distance D while keeping the same ID will lead to a smaller target, and hence increasing the difficulty of the selection while still considering the same ID. 
    On the one hand, this might reveal how distance impacts pointing performance.
    On the other hand, regression analysis with Fitts' law might not make sense since we try to model selection times with IDs not reflecting the real motor task.
    \item The same applies when varying amplitude A with linear measures. 
    Indeed, a large amplitude on large flat displays will lead to a smaller (and distorted) target toward the edge of the display.
    Thus, varying A has a direct confounding effect on W with flat displays. On the other hand, with angular values, even if the target does not have the same visual aspect depending on the display, the motor pointing movement will be the same.
\end{itemize}

\update{Previous studies comparing linear and rotational fitts law formulas found conflicting results. 
Kopper et al. \cite{kopper2010human,kopper2011understanding} showed that angular measurements model distal pointing better than linear measurements on a flat display. 
However, Shoemaker et al. \cite{shoemaker2012two} and Janzen et al. \cite{janzen2016modeling} showed their version of linear Fitts law performs better than the angular Fitts law. 
Due to the divide in previous literature, we compare both linear and angular measures in our study.}

\subsection{Distance Between Users and Displays}
Interaction distance is defined by the distance from the user perpendicular to the center of the display \cite{tao2021freehand, janzen2016modeling, shi2022pointing}.
Prior works reveal that users pointing performance varies based on the interaction distance \cite{kovacs2008perceptual, hourcade2012small, kopper2010human}. 
as it has a direct correlation with the target width in the display space, which further impacts the target selection time \cite{kovacs2008perceptual, hourcade2012small}: the further away from a display, the smaller the target.
To alleviate this confounding effect, Kopper et al. \cite{kopper2010human} integrated the interaction distance by using angular measurements of target width and amplitude. 
They showed that angular measures are more accurate for distal pointing tasks than linear measures.
Tao et al. \cite{tao2021freehand} evaluated the effects of interaction distance on pointing performance where they found that linear interaction distance significantly affected pointing accuracy and workload.
However, using angular measures is not yet commonly accepted in the HCI community \cite{shoemaker2012two, janzen2016modeling}.
We kept the distance between users and displays fixed in our study, but we plan to delve deeper into this aspect in the future.

\subsection{Curved-Display Specific Factors}
Curved displays, that come with concave viewing surfaces, are commonly determined by a curvature value in millimeters \update{\cite{kyung2021curved, lee2016factors}} followed by the letter "R", representing how curved the display is (e.g., 2000R refers to a screen having a radius of 2000 millimeters or 2 meters - see Figure \ref{all5displayTypes}).
Since previous work was limited by technological constraints, very few curvature values have been explored so far (e.g., 50" display with display curvature 400R, 600R, 1200R, and flat in \cite{kyung2021curved} and 55" with display curvature 5000R and flat in \cite{ahn2014research}).


\subsection{Exploration Summary}

We listed 2 factors - namely IDs (based on W and A), and curvature - and several Fitts law variations.
However, previous work does not allow us to clearly fix some factors (i.e., A and W with linear or angular units) or which model to actually use to draw conclusions about our exploration.

We hence consider and compare linear and angular measures for both (i) the definition of experimental condition, and (ii) the model accuracy as to not overlook any potential bias one measure unit would have on our conclusions. We compare pointing performances with three curved displays and a flat display using both units and the three Fitts law variations.

\subsection{Goals}
Our study has three main goals (G1, G2, and G3):
\begin{itemize}
    \item G1) To establish which measure unit (linear or angular) to use to define experimental conditions captured by Fitts law, a human motor behavior model.
    \item G2) To establish which model variations (standard, two-parts, or peephole) to use to explore pointing performances with different curvatures.
    \item G3) To identify users' pointing performance on large virtual displays for different display curvatures.
\end{itemize}

%% file: 6_exp-protocol.tex
\section{User Study}

\subsection{Display Curvature}
We designed 3 curved displays and compared them to a flat display (with $\infty$ R display curvature) as a baseline (Figure \ref{all5displayTypes}). 
For curved display, we included 2000R, 4000R, and 6000R curvatures where the numbers such as 2000 refer to screens having a display with a radius of 2000mm or 2m.
The display height was set to 3m. 


\subsection{Target Amplitude}
\begin{figure}[h]
	\centering
	\includegraphics[width=1.0\columnwidth]{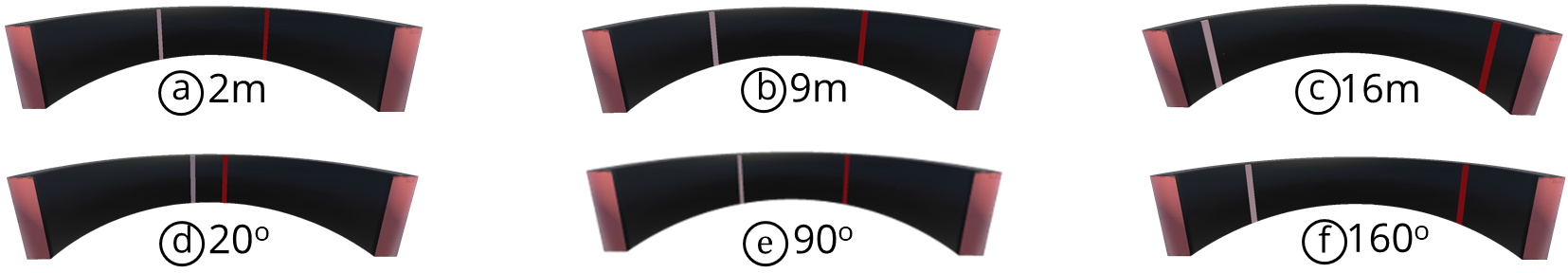}
	\caption{Linear target amplitude (a) 2m, (b) 9m, (c) 16m and  angular target amplitude (d) 20\degree, (e) 90\degree, (f) 160\degree}
	\label{fig:targetAmplitude}
\end{figure}

Based on prior work, we investigated both linear \cite{shoemaker2012two, janzen2016modeling, mackenzie1992extending, soukoreff2004towards} and angular target amplitudes \cite{kondraske1994angular, kopper2010human, kopper2011understanding}. 
We used 2m, 9m, and 16m as the linear target amplitudes for the linear condition (Figure \ref{fig:targetAmplitude}a-c).
\update{The corresponding angular amplitudes are calculated using Kopper et al.'s formula (\cite{kopper2010human}, equation 4) and are denoted as $\alpha$ (shown in appendix Table 4).}
The linear amplitude of 16m was placed on the extended circular region of the 2000R while maintaining the display curvature. 


For angular target amplitudes, we used 20\degree, 90\degree, and 160\degree (Figure \ref{fig:targetAmplitude}d-f) where the targets were placed from the user's standing position as suggested in \cite{kopper2010human}. \update{The angular amplitudes and their corresponding linear amplitudes (A) are listed in appendix Table 4.}

\subsection{Target Width}
\begin{figure}[h]
	\centering
	\includegraphics[width=1.0\columnwidth]{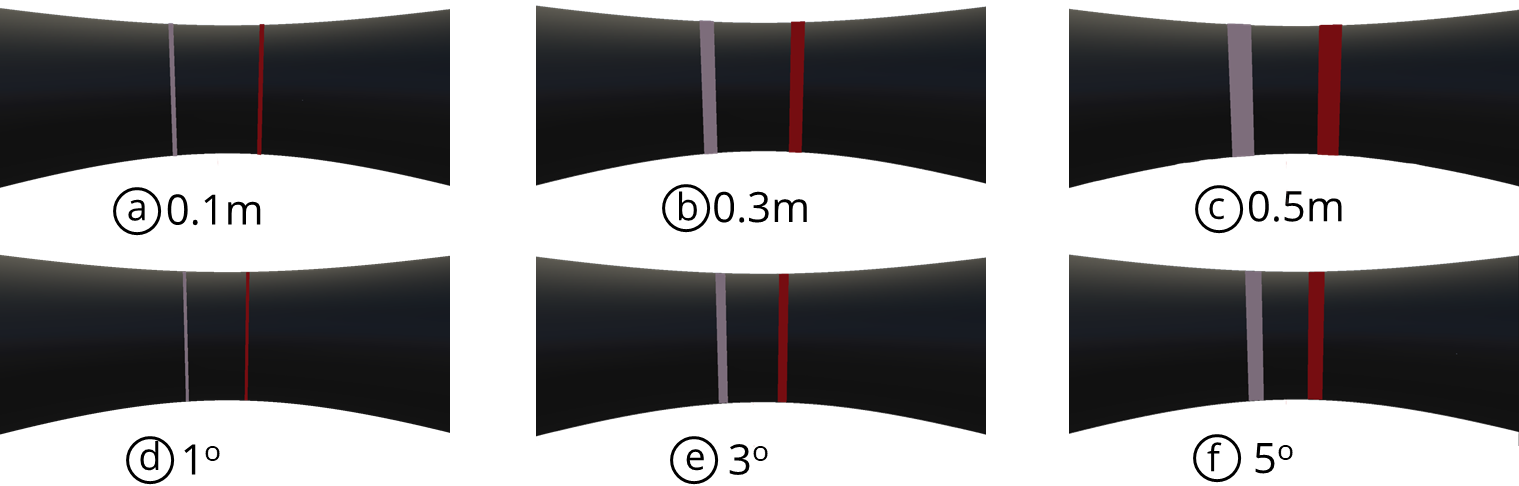}
	\caption{Linear target width (a) 0.1m, (b) 0.3m, (c) 0.5m and Angular target width (d) 1\degree, (e) 3\degree, (f) 5\degree} 
	\label{fig:targetWidth} 
\end{figure}

For linear target width, we used 0.1m, 0.3m, and 0.5m width. 
\update{We then used Kopper et al.'s formula (\cite{kopper2010human}, equation 5) to report the angular values ($\omega$) of the target widths for each curved display (Appendix Table 4).
For the angular widths, we consider 1\degree, 3\degree, and 5\degree  (Figure \ref{fig:targetWidth}d-f) and their corresponding linear values are denoted as (W).} 


\subsection{Interaction Distance}
 For pointing on large displays, interaction distance for prior research varied from 1m to 4m \cite{tao2021freehand, kopper2010human, jiang2006direct, lischke2016magic}. In our study, participants will be positioned at a fixed distance, i.e., 4 m from the screen (Figure \ref{all5displayTypes}e), on the middle axis going through the screen center. 
This will ensure the least deviation of the users' position from the center of each semi-circular display: toward the back for 2000R, at the center for 4000R, and toward the front for 6000R.
Further study can investigate how users’ pointing performance differs due to the change in their relative position from the center of the large curved displays.

\subsection{Participants and Apparatus}
We recruited 12 participants (8 males, \update{4 females}, aged between 21 and 33 years, avg. 25.92, SD. 3.26, all right-handed) from a local university using on-campus and online flyers. 
Only four participants had at least one-year of prior experience using an HMD for VR applications. 
Each participant received \$15 as compensation for participating in the study. 

We created a virtual environment with large displays using Unity \cite{unity}.
We used the Oculus Quest 2 \cite{meta}, with a horizontal field-of-view of 90\degree, and the Quest 2 right controller to point to the virtual displays. 
Firebase Realtime Database \cite{google} was used for logging data during the study.  

\subsection{Procedure}
We asked participants to wear the Quest 2 HMD and use a controller on their dominant hand to perform a 1D Fitts task.
We used vertical strip targets (i.e., bars) on the virtual displays (Figure \ref{fig:targetWidth}).
The bars had the same height as the virtual display (i.e., 3 m).
Participants were standing 4m away from the center of the screen - facing toward the large display.
Participants could use the hand-held controller to manipulate a ray originating from the controller.
They were instructed to move the endpoint of the ray on top of a target and press the device's trigger to validate a selection.
The `start' target was always located at the center of the virtual display.
The `start' target disappeared after a successful selection, and the system revealed two new targets (red and grey) on the virtual display for the reciprocal selection task.
\update{Participants had to complete 10 successful selections (altering between the two targets) after which the targets disappeared and "start" reappeared.}
If a participant selected outside the target (i.e., pressed the controller button when the ray was outside the targets), the trial was marked as an error trial, and the application continued to the next trial. 
The error trials were discarded and then re-queued into the remaining trial pool.
During the study, we instructed participants to point to the targets quickly while maintaining accuracy.

\subsection{Design}

We used a within-subject design with three independent variables: display curvature (Flat, 2000R, 4000R and 6000R), target amplitude (20$^{\circ} $, 90$^{\circ}$ and  160$^{\circ}$ for angular measurements and 2m, 9m and 16m for linear measurements), target width (1$^{\circ} $, 3$^{\circ} $ and 5$^{\circ}$ for angular measurements and 0.1m, 0.3m, and 0.5m for linear measurements). 
Note that for the linear target amplitude, we only used the linear target widths and for angular target amplitude, we only used the angular target widths.
These values lead to nine \textit{Index of Difficulty (ID)} ranging from 2.32 to 7.33 bits which were the same for both linear and angular measurements when considering the standard Shanon ID.
Measure types (linear vs. angular measures) were counterbalanced across participants.
\update{For each measure type, Curvature (2000R, 4000R, 6000R, and Flat) were separately counterbalanced across participants.
IDs were randomized.}
For each measure type, there were 360 trials per participant (4 display curvatures $\times$ 3 target amplitudes $\times$ 3 target widths $\times$ 10 repetitions) for a total of 4320 trials. 
Participants were given 15 trials to help them familiarize themselves with the study conditions.
Additionally, they were asked to take breaks to eliminate fatigue. 
The study lasted approximately 60 minutes.

\subsection{Measurement}
We recorded the movement time defined by the time to select a target, i.e., the time between two valid selections. 
\update{We define endpoint deviation (also known as spread of hits) \cite{fitts1954information, wobbrock2008error, wobbrock2011effects} as the distance between the target’s center and the selection point along the horizontal axis of the display surface.}
For the linear measures, we considered endpoint deviations in meters whereas for the angular measurements, we considered endpoint deviations in degrees.
In addition, we also logged the number of error trials that the participants made during the study.
We define error trial as the percentage of trials where the participants had made at least one selection outside the boundary of the target.
\update{We then calculated the error rate as the percentage of trials with errors over the total number of trials.}
We collected participants' demographic information, their subjective feedback, and ratings for each of the investigated factors. 

%% file: 7_exp-results.tex
\begin{figure*}[htbp]
	\centering
	\includegraphics[width=2.0\columnwidth]{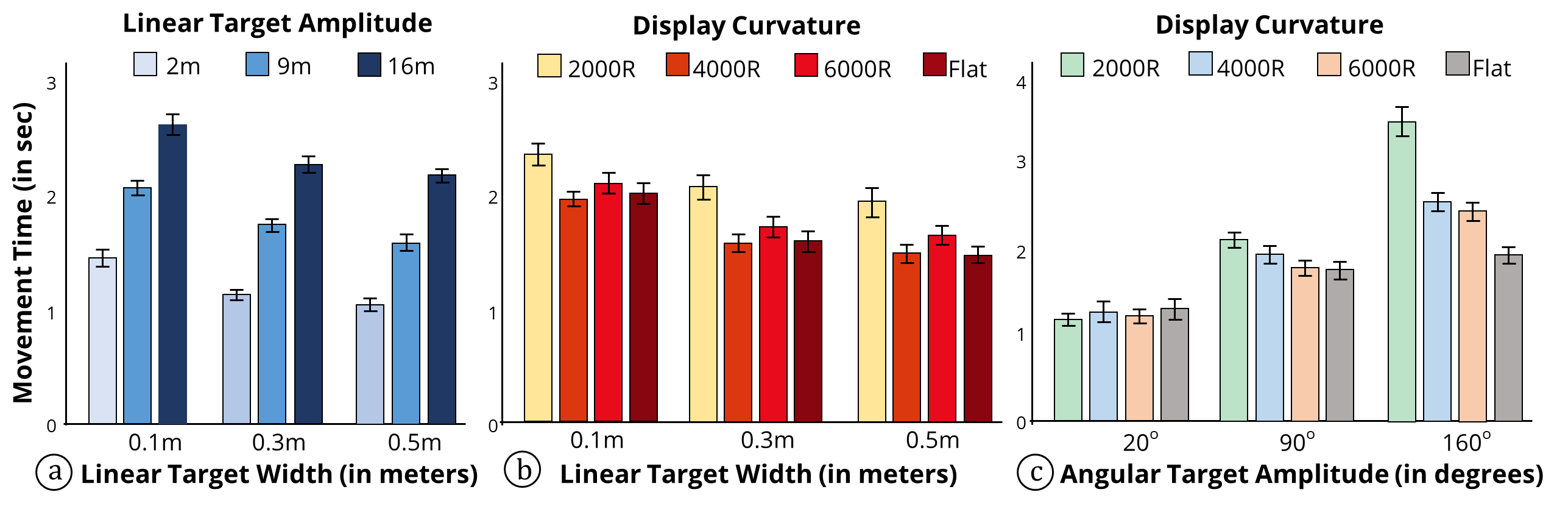}
	\caption{Movement time by: (a) linear target width for each linear target distance, (b) linear target width for each display curvature and (c) angular target amplitude for each display curvature. \update{Error bars represent 95\% confidence intervals.}}
	\label{fig:exp1-movementtime} 
\end{figure*}

\section{Study Results}

We removed movement times more than 3 standard deviations from the average. 
This led to 25 out of 4320 (0.58\%) trials being removed from the angular measurement data and 30 out of 4320 (0.69\%) trials being removed from the linear measurement data.

\update{We initially counterbalanced both measures (2 levels) \& curvatures (4 levels) with a Latin-square design. We have 2×4= 8 orderings and only 12 participants, hence not a fully counter-balanced design.
We ensure that six participants started with the linear condition, and six others started with the angular condition, and that we had three participants for each one of the four ordering of the curvature conditions.
Even if balanced, this design is not fully counter-balanced.
We hence first confirm that there is no confounding effect of the ordering itself on movement time using Mann-Whitney's U tests and \emph{ordering} as a between-subject factor for both linear and angular conditions (all $p > 0.05$).
}

\update{We next apply a log transformation to normalize our skewed data.}
We \update{then} analyze movement time and endpoint deviation with repeated measures ANOVA on log-transformed data, and post-hoc pairwise comparisons with Bonferroni corrections.
In case of sphericity violation, we report Greenhouse-Geisser corrected p-values and degrees of freedom.

\begin{table*}[]
\caption{ANOVA main effects and interactions for movement
time (MT).} 
\begin{center}
 \begin{adjustbox}{width=0.8\textwidth}
\begin{tabular}{|c|cc|cc|}
\hline
Factor(s) & \multicolumn{2}{c|}{Linear} & \multicolumn{2}{c|}{Angular} \\ \cline{2-5} 
 & \multicolumn{1}{c|}{Sig?} & ANOVA results & \multicolumn{1}{c|}{Sig?} & ANOVA results \\ \hline
Display & \multicolumn{1}{c|}{*} & $F_{1.75 , 19.27}  = 11.8$, $p<0.001$, $\eta^2$=0.52 & \multicolumn{1}{c|}{*} & $F_{3, 33}  = 18.12$, $p<0.0001$, $\eta^2$=0.62 \\ \hline
Amplitude & \multicolumn{1}{c|}{*} & $F_{2 , 22}  = 334.76$, $p<0.0001$, $\eta^2$=0.97 & \multicolumn{1}{c|}{*} & $F_{2 , 22}  = 271.19$, $p<0.0001$, $\eta^2$=0.96 \\ \hline
Width & \multicolumn{1}{c|}{*} & $F_{2 , 22}  = 66.34$, $p<0.0001$, $\eta^2$=0.86 & \multicolumn{1}{c|}{*} & $F_{2 , 22}  = 218.57$, $p<0.0001$, $\eta^2$=0.95 \\ \hline
Display: Amplitude & \multicolumn{1}{c|}{} & $F_{6 , 66 }=2.09$, $p=0.07$ & \multicolumn{1}{c|}{*} & $F_{6 , 66 }=23.87$, $p<0.0001$, $\eta^2=0.68$ \\ \hline
Display: Width & \multicolumn{1}{c|}{*} & $F_{6 , 66 }=3.29$, $p<0.01$, $\eta^2=0.23$ & \multicolumn{1}{c|}{} & $F_{6 , 66 }=0.89$, $p=0.51$ \\ \hline
Amplitude: Width & \multicolumn{1}{c|}{*} & $F_{4,44}=6.07$, $p<0.001$, $\eta^2=0.36$ & \multicolumn{1}{c|}{} & $F_{4,44}=0.83$, $p=0.52$ \\ \hline
Display: Amplitude: Width & \multicolumn{1}{c|}{} & $F_{5.22,57.37}=1.06$, $p=0.39$ & \multicolumn{1}{c|}{} & $F_{4.07,44.82}=1.22$, $p=0.31$ \\ \hline
\end{tabular}
 \end{adjustbox}
 \end{center}
\end{table*}


\subsection{Movement time}

\subsubsection{Linear Measurement}
In terms of movement time for linear measurements, we found significant main effects for the independent variables \textit{Display Curvature} ($F_{1.75, 19.27} = 11.8$, $p < 0.001$, $\eta^2$ = 0.52), \textit{Amplitude} ($F_{2, 22} = 334.76$, $p < 0.001$, $\eta^2$ = 0.97), and \textit{Width} ($F_{2, 22}  = 66.34$, $p < 0.001$, $\eta^2$ = 0.86). The mean movement time is 2.1s (\update{95\% confidence interval, CI:} [1.95, 2.25]) for 2000R, 1.65s (\update{CI:} [1.54, 1.75]) for 4000R, 1.81s (\update{CI:} [1.69, 1.94]) for 6000R, and 1.66s (\update{CI:} [1.55, 1.77]) for Flat. Post-hoc pairwise comparisons reveal that pointing times on 4000R, 6000R, and Flat are significantly faster than 2000R (all $p < 0.001$). In addition, pointing on 4000R and Flat is significantly faster than 6000R (all $p < 0.001$).
We found significant interactions of \textit{Display Curvature} $\times$ \textit{Width} ($F_{6, 66} = 3.29$, $p < 0.01$, $\eta^2 = 0.23$) (Figure \ref{fig:exp1-movementtime}b) and \textit{Amplitude} $\times$ \textit{Width} ($F_{4, 44} = 6.07$, $p<0.001$, $\eta^2=0.36$) (Figure \ref{fig:exp1-movementtime}a). There was no significant interaction effect between \textit{Display Curvature} $\times$ \textit{Amplitude} ($F_{6 , 66 }=2.09$, $p=0.07$).

\subsubsection{Angular Measurement}
For angular measurements, we found significant main effects for the independent variables \textit{Display Curvature} ($F_{3, 33}  = 18.12$, $p < 0.001$, $\eta^2$ = 0.62), \textit{Amplitude} ($F_{2, 22}  = 271.19$, $p < 0.001$, $\eta^2$ = 0.96), and \textit{Width} ($F_{2, 22}  = 218.57$, $p < 0.001$, $\eta^2$ = 0.95) on movement time. The mean movement time is 2.29s (\update{CI:} [2.05, 2.54]) for 2000R, 1.92s (\update{CI:} [1.77, 2.07]) for 4000R, 1.82s (\update{CI:} [1.68, 1.95]) for 6000R, 1.67s (\update{CI:} [1.55, 1.77]) for Flat. Post-hoc pairwise comparisons reveal that pointing on 4000R, 6000R, and Flat is significantly faster than 2000R (all $p < 0.001$ ). In addition, pointing on Flat is significantly faster than 2000R and 4000R (all $p < 0.001$).
We found significant interactions of \textit{Display Curvature} $\times$ \textit{Amplitude} ($F_{6, 66} = 23.87$, $p < 0.001$, $\eta^2 = 0.68$) (Figure \ref{fig:exp1-movementtime}c). There was no significant interaction effect between \textit{Display Curvature} $\times$ \textit{Width} ($F_{6, 66} = 0.89$, $p = 0.51$) and \textit{Amplitude} $\times$ \textit{Width} ($F_{4, 44} = 0.83$, $p = 0.51$).


\subsection{Endpoint Deviation}
\subsubsection{Linear Measurement}
For linear measurements, we found significant main effects for \textit{Display Curvature} ($F_{3, 33} = 22.3$, $p < 0.001$, $\eta^2$ = 0.67), \textit{Amplitude} ($F_{2, 22}  = 99.63$, $p < 0.001$, $\eta^2$ = 0.90), and \textit{Width} ($F_{1.27, 14.01}  = 172.16$, $p < 0.001$, $\eta^2$ = 0.94) on endpoint deviation. 
The mean endpoint deviation is 0.06m (\update{CI:} [0.06, 0.06]) for 2000R, 0.06m (\update{CI:} [0.05, 0.06]) for 4000R, 0.07m (\update{CI:} [0.07, 0.08]) for 6000R and 0.07m (\update{CI:} [0.06, 0.07]) for Flat. 
Post-hoc pairwise comparisons reveal that pointing on 4000R and 2000R have significantly less endpoint deviation than on 6000R and Flat (all $p < 0.001$).
We found significant interactions of \textit{Display Curvature} $\times$ \textit{Amplitude} ($F_{6, 66} = 21.45$, $p < 0.001$, $\eta^2 = 0.66$), \textit{Display Curvature} $\times$ \textit{Width} ($F_{6, 66} = 7.81$, $p < 0.001$, $\eta^2 = 0.42$)  and \textit{Amplitude} $\times$ \textit{Width} ($F_{4, 44} = 12.84$, $p < 0.001$, $\eta^2 = 0.54$) on endpoint deviation. There was also a three-way interaction among \textit{Display Curvature} $\times$ \textit{Amplitude} $\times$ \textit{Width} ($F_{5.61, 61.74} = 6.18$, $p < 0.001$, $\eta^2 = 0.36$) on endpoint deviation.


\subsubsection{Angular Measurement}
For angular measurements, we found significant main effects for \textit{Display Curvature} ($F_{1.21, 13.35} = 59.11$, $p < 0.001$, $\eta^2$ = 0.84), \textit{Amplitude} ($F_{2 , 22} = 169.8$, $p < 0.001$, $\eta^2$ = 0.94), and \textit{Width} ($F_{1.11, 12.26}  = 967.52$, $p < 0.001$, $\eta^2$ = 0.99) on endpoint deviation. The mean endpoint deviation is 1.03\degree (\update{CI:} [0.91 , 1.15]) for 2000R, 0.97\degree (\update{CI:} [0.86 , 1.08]) for 4000R, 0.73\degree (\update{CI:} [0.68, 0.78]) for 6000R and 0.59\degree (\update{CI:} [0.53, 0.64]) for Flat. Post-hoc pairwise comparisons revealed that pointing on Flat and 6000R is significantly more accurate than on 4000R and 2000R (all $p < 0.001$). Moreover, Flat is significantly more accurate than 6000R (all $p < 0.001$).  
We found significant interaction effects of \textit{Display Curvature} $\times$ \textit{Amplitude} ($F_{1.74, 19.1} = 100.21$, $p < 0.001$, $\eta^2 = 0.90$), \textit{Display Curvature} $\times$ \textit{Width} ($F_{2.62, 28.82} = 63.71$, $p < 0.001$, $\eta^2 = 0.85$)  and \textit{Amplitude} $\times$ \textit{Width} ($F_{4, 44} = 165.89$, $p < 0.001$, $\eta^2 = 0.94$) on endpoint deviation. There was also a three-way interaction among \textit{Display Curvature} $\times$ \textit{Amplitude} $\times$ \textit{Width} ($F_{12, 132} = 30.76$, $p < 0.001$, $\eta^2 = 0.74$).

\subsection{Error Rate}
\subsubsection{Linear Measurement}
For the Linear measurement, we found error rate of 1.95\% for the 2000R display, 1.82\% for the 4000R display, 3.18\% for the 6000R display, and 3.96\% for the Flat display. One-way ANOVA showed that there is a significant difference in the error rate for the different display curvatures ($F_{3, 4316} = 16.91$, $p < 0.001$, $\eta^2 = 0.01$). Post-hoc tests with Bonferroni corrections showed that 2000R and 4000R display has significantly lower rates of error trials (all $p < 0.01$) than 6000R and Flat displays.
\subsubsection{Angular Measurement}
For the Angular measurement, the error rate is 4.78\% for 2000R, 3.78\% for the 4000R, 2.35\% for the 6000R, and 1.76\% for the Flat display. One-way ANOVA showed that there is a significant difference in the error trial rate among the different display curvatures ($F_{3, 4316} = 13.00$, $p < 0.001$, $\eta^2 = 0.01$). Post-hoc tests with Bonferroni corrections showed that Flat and 6000R has significantly fewer error trials (all $p < 0.001$ ) than the 2000R display. Additionally, Flat has significantly fewer error trials than the 4000R display ($p < 0.001$ ).


\subsection{Throughput Analysis}
Prior work used Throughput (TP) as a fundamental metric while quantifying input system performance \cite{fitts1954information, iso20009241, zhai2004speed, mackenzie1992fitts}. In Fitts Law, movement time and difficulty index change due to width and amplitude changes. 
As throughput considers both speed (movement time) and accuracy (endpoint deviation), it remains the same for a particular interaction or input device.   
To remove subjective speed-accuracy biases \cite{zhai2004speed} for pointing, we use  Crossman’s correction \cite{crossman1957speed} to find the effective Index of Difficulty, $ID_e$. Here, $ID_e =\log _2(\frac{A_e}{W_e}+1)$. Here, $W_e$ denotes the effective width based on a standard deviation of endpoints as $W_e = 4.133\times SD$ \update {and $A_e$ is the mean amplitude between the start position and the end position of the pointer for a certain amplitude condition.}
Following the recommendations of Wobbrock et al. \cite{wobbrock2011effects}, we used the means-of-means approach \cite{soukoreff2004towards, iso20009241} for throughput computations. 
For each display curvature, we compute the throughput TP as $TP=\frac{1}{N}\sum_{i=1}^N\left(\frac{ID_{e_i}}{M T_i}\right)$, \update{where $N= 9$ (multiplication of the cardinality of amplitude and width conditions: $card(A_e) \times card(W_e)$).}

\begin{figure}[htbp]
	\centering
	\includegraphics[width=0.95\columnwidth]{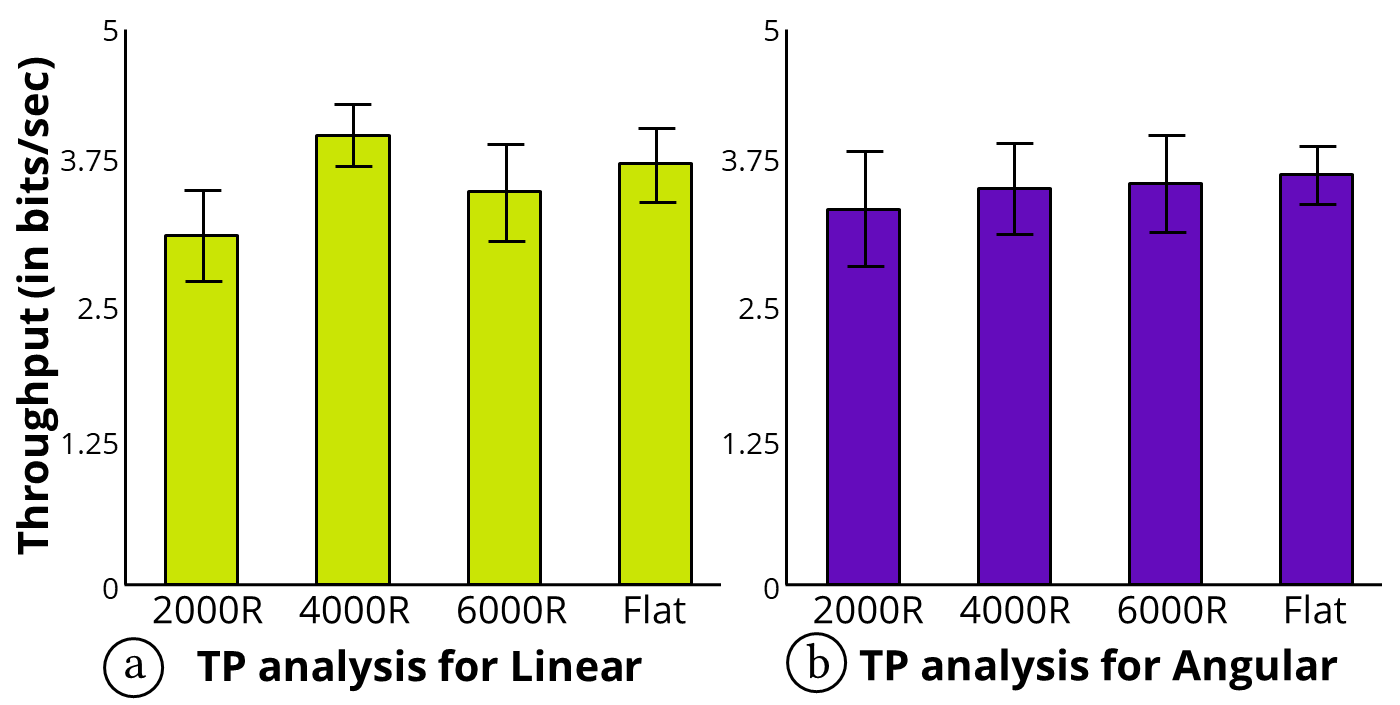}
	\caption{Throughput by display curvature for: (a) linear, (b) angular measurement. \update{Error bars represent 95\% confidence intervals.}}
	\label{fig:exp1-ThroughputandMT} 
\end{figure}

\subsubsection{Linear Measurement}
For the linear data, the mean throughput ($TP$) is 3.10 bits/s (\update{CI:} [2.75, 3.44]) for 2000R, 3.98 bits/s (\update{CI:} [3.75, 4.22]) for 4000R, 3.48 bits/s (\update{CI:} [3.12, 3.85]]) for 6000R and 3.73 bits/s (\update{CI:} [3.46, 4.00]) for Flat (Figure \ref{fig:exp1-ThroughputandMT}a). 
A one-way ANOVA \cite{soukoreff2004towards} showed that \emph{Display Curvature} ($F_{3, 9} = 5.68$, $p < 0.01$, $\eta^2 = 0.35$) has significant effect on throughput. 
Posthoc tests revealed only the 4000R display had higher significantly higher throughput than the 2000R display ($p<0.01$). 

\subsubsection{Angular Measurement}
For the angular data, we used angular amplitude and angular values of endpoints to calculate $W_e$. 
The throughput ($TP$) is 3.32 bits/s (\update{CI:} [2.89, 3.76]]) for 2000R, 3.51 bits/s (\update{CI:} [3.17, 3.85]) for 4000R, 3.56 bits/s (\update{CI:} [3.18, 3.93]) for 6000R and 3.64 bits/s (\update{CI:} [3.42, 3.86]) for Flat (Figure \ref{fig:exp1-ThroughputandMT}b). 
We did not find any significant effect of \emph{Display Curvature} ($F_{3, 9} = 1.16$, $p = 0.34$) on throughput.

\begin{figure*}[h]
	\centering
	\includegraphics[width=2.0\columnwidth]{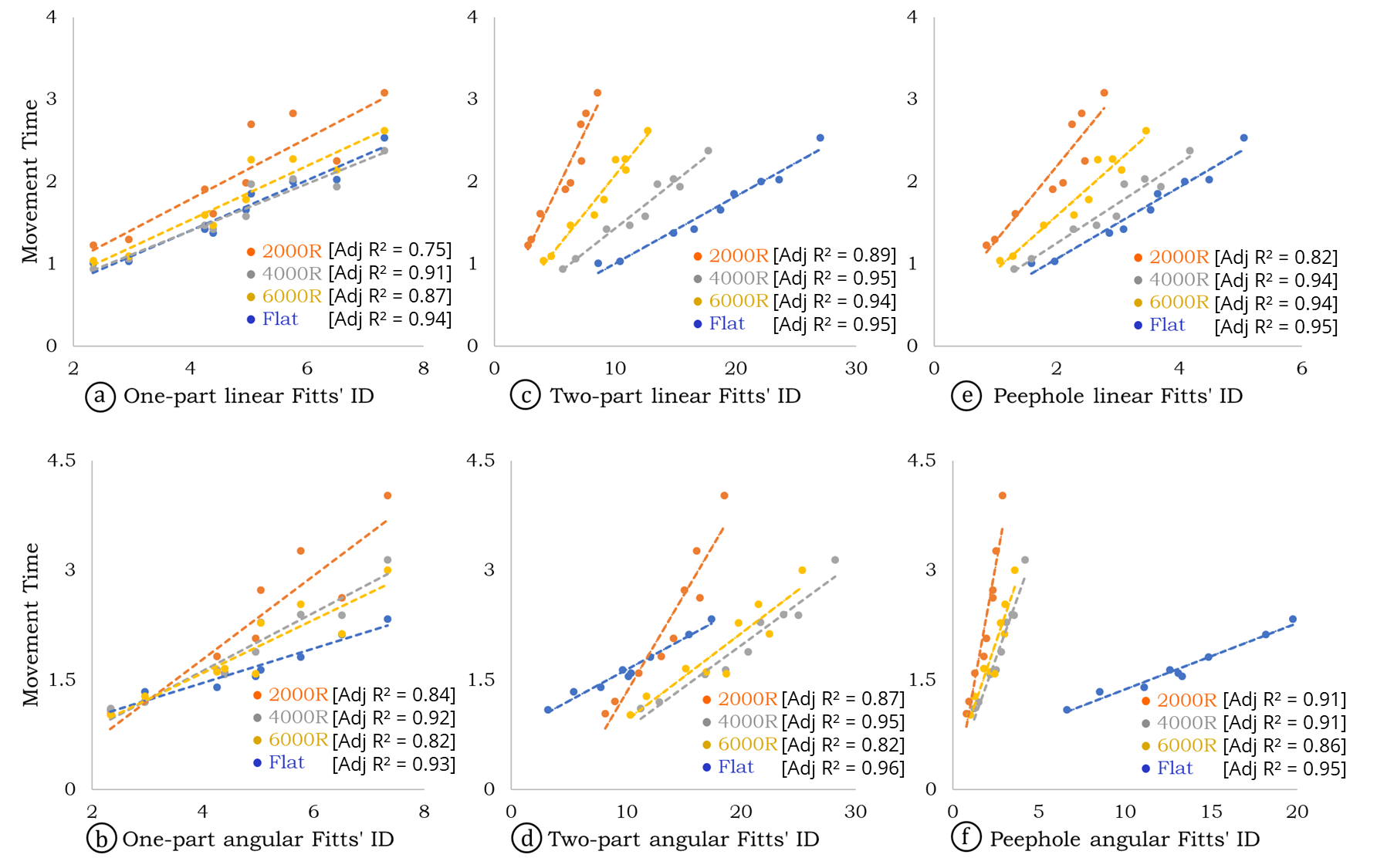}
	\caption{Fitts law model regression line for each display curvature: One-part model with (a) linear and (b) angular measurements, Two-part regression with (c) linear and (d) angular measurements, Peephole model with (e) linear and (f) angular measurements.}
	\label{fig:exp1-overallFitts} 
\end{figure*}

\subsection{Preference Score}

The participants filled out a questionnaire providing feedback on their mental demand, physical demand, temporal demand, performance, effort, frustration, and preference on a 5-point Likert scale for \textit{Display Curvature} and \textit{Measurement}. 

\subsubsection{Linear Measurement}
A Friedman test on the linear measurement data shows that there is a significant difference for all seven criteria: mental demand ($\chi^2 (3, N=12) = 15.97, p < 0.001$), physical demand ($\chi^2 (3, N=12) = 22.36, p < 0.001$), temporal demand ($\chi^2 (3, N=12) = 17.07, p < 0.001$), performance ($\chi^2 (3, N=12) = 9.82, p < 0.05$), effort ($\chi^2 (3, N=12) = 16.03, p < 0.001$), frustration ($\chi^2 (3, N=12) = 18.45, p < 0.001$) and preference ($\chi^2 (3, N=12) = 18.71, p < 0.001$)) across the four display curvatures. A Wilcoxon signed-rank (Bonferroni: $\alpha$-levels from 0.05 to 0.008) test shows that 4000R display ranks better than 2000R across all criteria. 6000R display has better ratings for Fatigue, Physical Demand, and Temporal Demand than the 2000R display. The Flat display has better ratings for Fatigue and Physical Demand than the 2000R display.   


\subsubsection{Angular Measurement}
A Friedman test on the Angular Measurement data shows that there is a significant difference for only Effort ($\chi^2 (3, N=12) = 16.07, p < 0.001$) and Frustration ($\chi^2 (3, N=12) = 16.18, p < 0.001$) between the display curvatures. A Wilcoxon signed-rank test (Bonferroni: $\alpha$-levels from 0.05 to 0.008) shows that the Flat display has better ratings than 2000R for Effort and Frustration. Additionally, the 4000R display has better ratings for Frustration than the 2000R display. 


Users perform faster selection times with the flat display than with the curved displays while curved displays have similar to higher throughput than flat ones, for both linear and angular conditions.
We next continue our analysis with linear regressions and three Fitts law variations to get a better understanding of the effect a flat display has on movement times.

\subsection{Fitts' Law Regression Lines}
Each point in the regression analysis represents an average movement time for a particular condition based on target width, target amplitude, and display curvature for both linear and angular measurements. 
Beside standard fits measured via  Adj $R^2$ (Adjusted $R^2$ value), we also evaluate how the models perform compared to each other via F-tests and p-values. Appendix Table 3 shows the performance metrics (i.e., $R^2$, Adj $R^2$, F-statistics score, AIC \cite{akaike1974new, batmaz2020effect}, and BIC \cite{schwarz1978estimating} values) for each of the Fitts law models.

\subsubsection{One-part}
Overall, we found that the one-part model (Eq. \ref{eqn_Fittslaw_popular}) provides a fit of Adj $R^2=0.88$ for linear measurements whereas it offers a higher Adj $R^2 = 0.89$ for the angular measurements (Appendix Table 3). This also indicates that the model could accurately explain about 88\% and 89\% of the data points with linear and angular measures, respectively.

\textbf{Linear}: The one-part model predicts movement time accurately with all display curvatures (Adj $R^2$ ranging from 0.75 with 2000R to 0.94 for Flat) (Figure \ref{fig:exp1-overallFitts}a).
Interestingly, with this model, all predictions have the same slope (b $\approx$ 0.25), indicating that all display curvatures exhibit the same tolerance to an increase in task difficulty.

\textbf{Angular}: The one-part model predicts angular conditions slightly better than linear ones (Adj $R^2$ from 0.82 with 6000R to 0.93 with Flat) (Figure \ref{fig:exp1-overallFitts}b).
Interestingly, the higher the display curvature, the lower the movement time and slope: ($MT_{FlatR} < MT_{6000R} \leq MT_{4000R} < MT_{2000R}$).
Note that 4000R and 6000R lead to very similar regression lines - which indicates that pointing tasks controlled by angular values indeed lead to similar movement times. we hypothesize that the strong difference with 2000R resides in the fact that large amplitudes (i.e., large IDs) are correlated to targets placed farther away from the user.

\textbf{Summary}: Overall, the one-part model can predict movement times slightly better with angular measurements rather than linear measurements with varying display curvatures based on Adj $R^2$ values and F-statistic values (Appendix Table 3). With this model, the Flat display's advantage resides in its resistance to an increase in difficulty (i.e., low slope of the regression line).

\subsubsection{Two-part}
The two-part Fitts' model is essentially multiple linear regression that considers the effect of amplitude and width separately. Overall, the two-part model (Eq \ref{eqn_Fittslaw_twopart}) provides a fit of Adj $R^2=0.94$ for linear measurements and Adj $R^2=0.90$ for angular measurements (Appendix Table 3). 
For both linear and angular conditions, Fitts IDs for each display curvature are different due to the $k$ parameter in equation \ref{eqn_Fittslaw_twopart}.

\textbf{Linear}: The two-part model predicts movement time accurately with all display curvatures (Adj $R^2$ ranging from 0.89 with 2000R to 0.95 for Flat) (Figure \ref{fig:exp1-overallFitts}c).
Here, the slope decreases as the curvature increases ($Slope_{2000R} > Slope_{4000R} > Slope_{6000R} > Slope_{Flat}$).
In addition, IDs span a larger range for larger curvatures.
Thus, the flat display (i) is more resistant to task difficulty than curved displays while (ii) covering a wider range of selection difficulty.

\textbf{Angular}: The two-part model predicts angular conditions accurately with all curvature (Adj $R^2$ from 0.82 with 6000R to 0.96 with Flat) (Figure \ref{fig:exp1-overallFitts}d).
As for the one-part model, (i) regression lines of 4000R and 6000R reveal a similar behavior (slopes and IDs) and (ii) the regression line for 2000R reveals a high slope due to higher IDs leading to long movement times.
The regression line of the flat display has the lower slope of all lines.
Contrary to the one-part model, two-part model considers that IDs for the flat display are lower than with other displays and cover a smaller range of values.

\textbf{Summary}: 
With this model, the Flat display's advantage resides in two potential reasons: a resistance to an increase of task difficulty (lower slope, similar to the one-part model), and a range of IDs lower than with the other displays.

\subsubsection{Peephole}

The peephole variation (Eq \ref{eqn_Fittslaw_Peephole}) depends on \textit{Screen Size}, target \textit{Amplitude} and target \textit{Width}. 
We consider a display size $S$ of 90\degree - the horizontal field-of-view of the Oculus Quest 2 headset.
To compute IDs, we need to have unitless terms in the logarithm parts, i.e., $W/S$ and $A/S$.
We hence use angular measures for both W and A. For linear conditions, we use their angular computed counter-parts (Appendix Table 4).
As for the two-part model, the extra parameter $n$ leads to different IDs depending on the display curvature condition. 

Overall, the peephole variation (Eq \ref{eqn_Fittslaw_Peephole}) provides a fit of Adj $R^2=0.91$ for linear and Adj $R^2=0.93$ angular conditions.

\textbf{Linear}: The peephole model accurately predicts movement times for all display curvatures (from Adj $R^2=0.82$ for 2000R to Adj $R^2=0.95$ for Flat) (Figure \ref{fig:exp1-overallFitts}e). Results are similar to the two-part model: the slope decreases with an increase of curvature, and Flat IDs cover a wider range of values than with curved displays.

\textbf{Angular}: The peephole model accurately predicts angular conditions with all curvatures (Adj $R^2$ from 0.86 with 6000R to 0.95 with Flat) (Figure \ref{fig:exp1-overallFitts}f).
However, the model reveals a clear contrast between curved displays and the Flat display: regression lines of curved displays constitute a cluster with low IDs and high slopes, while the regression line of Flat covers a wider range of higher IDs.

\textbf{Summary}: 
(Appendix Table 3) 
With this model, the Flat display's advantage resides in its resistance to difficulty and its ability to leverage a wide range of high IDs. However, the peephole model with angular values reveals a stronger differentiation between Flat and curved displays than the other model variations.

\begin{table*}[]
\begin{center}
\caption{AIC and BIC comparisons for One-part, Two-part and Peephole model.}
\label{tab:AICBIC}
 \begin{adjustbox}{width=0.8\textwidth}
\begin{tabular}{|l|ll|lllllll|}
\hline
         & \multicolumn{2}{l|}{$\Delta_i AIC$}            & \multicolumn{7}{c|}{$\Delta BIC$}                                                                                                                                                                                \\ \hline
         & \multicolumn{1}{l|}{Linear} & Angular & \multicolumn{1}{l|}{}         & \multicolumn{3}{c|}{Linear}                                                                   & \multicolumn{3}{l|}{Angular}                                             \\ \hline
         & \multicolumn{1}{l|}{}       &         & \multicolumn{1}{l|}{}         & \multicolumn{1}{l|}{One-part} & \multicolumn{1}{l|}{Two-part} & \multicolumn{1}{l|}{Peephole} & \multicolumn{1}{l|}{One-part} & \multicolumn{1}{l|}{Two-part} & Peephole \\ \hline
One-part & \multicolumn{1}{l|}{8.42}   & 11.65   & \multicolumn{1}{l|}{One-part} & \multicolumn{1}{l|}{0}        & \multicolumn{1}{l|}{8.07}     & \multicolumn{1}{l|}{4.45}     & \multicolumn{1}{l|}{0}        & \multicolumn{1}{l|}{11.65}    & 2.29     \\ \hline
Two-part & \multicolumn{1}{l|}{0}      & 0       & \multicolumn{1}{l|}{Two-part} & \multicolumn{1}{l|}{8.07}     & \multicolumn{1}{l|}{0}        & \multicolumn{1}{l|}{3.62}     & \multicolumn{1}{l|}{11.65}    & \multicolumn{1}{l|}{0}        & 9.36     \\ \hline
Peephole & \multicolumn{1}{l|}{3.61}   & 9.35    & \multicolumn{1}{l|}{Peephole} & \multicolumn{1}{l|}{4.45}     & \multicolumn{1}{l|}{3.62}     & \multicolumn{1}{l|}{0}        & \multicolumn{1}{l|}{2.29}     & \multicolumn{1}{l|}{9.36}     & 0        \\ \hline
\end{tabular}
\end{adjustbox}
\end{center}
\end{table*}

\subsubsection{Models Comparison}

\textbf{The Akaike Information Criterion (AIC)} \cite{akaike1974new} and the \textbf{Bayesian Information Criterion (BIC)} \cite{schwarz1978estimating} are both metrics to compare between several regression models which considers the fitness of the model while penalizing models for added coefficients. The lower the score, the better the model.
\update{Considering the overall linear and angular measurement data for the each model, we compute $\Delta AIC_i$ = AIC of a $i^{th}$ model minus the lowest AIC ($AIC_{min}$) and $\Delta BIC$= Difference between BIC values of two models (see Table \ref{tab:AICBIC}). 
To evaluate if there are any significant differences between the models, we use  Burnham and Anderson \cite{burnham2004multimodel} criterion ($\Delta AIC_i<2$ = substantial evidence, $2<\Delta AIC_i<4$ = strong evidence, $4<\Delta AIC_i<7$ = less evidence, $\Delta AIC_i>10$ = no evidence) for AIC values  and Raftery \cite{raftery1995bayesian} criterion  ($\Delta BIC>10$ = very strong evidence, $6<\Delta BIC<10$ = strong evidence, $2<\Delta BIC<6$ = positive evidence, $0<\Delta BIC<2$ = no evidence) for BIC values.}

\update{The lowest AIC ($AIC_{min}$)  corresponds to the Two-Part model (linear: -10.46, angular: -11.16).}
\update{Overall, based on AIC differences, the two-part model (Eq. \ref{eqn_Fittslaw_twopart}) can predict movement times significantly better than the one-part model (linear: $\Delta AIC_i = 8.42$, angular: $\Delta AIC_i = 11.65$, less to no evidence for the one-part model) and peephole model (strong evidence for linear: $\Delta AIC_i = 3.61$, less to no evidence for angular: $\Delta AIC_i = 9.35$ for the peephole model). 
This is also supported by the differences between BIC values, revealing a strong evidence of better fit compared to the one-part model (linear: $\Delta BIC=8.07$, angular: $\Delta BIC=11.65$), and a positive evidence when compared to the peephole model (linear: $\Delta BIC=4.45$, angular: $\Delta BIC=2.29$).} 


%% file: 10_discussion.tex
\section{Discussion}
We interpret key insights from our results and discuss them for large displays in virtual reality. 


\subsection{Distal Pointing Performance Modeling}

\subsubsection{Linear versus Angular Measures}
Our results support the recommendation of Kopper et al. \cite{kopper2010human} and George Kondraske \cite{kondraske1994angular} to use angular measurements for experimental conditions to model VR pointing performance. 
Angular values (for amplitudes and widths) perform better than linear ones as angular values adapt the targets to the display curvature.
The advantage of using angular amplitude is that angular amplitude is consistent across different display curvatures. 
\update{Therefore, we suggest researchers to use angular values to better capture the actual motor movement performances for both (i) the definition of the experimental conditions and (ii) the result analyses.}

\subsubsection{Fitts' Law Variations}

Interestingly, all three models manage to capture our distal pointing tasks relatively accurately, thus showing that Fitts Law and its variations predict users pointing performance. 
However, we observe that the two parts and the peephole models have higher $R^2$ values than the standard Fitts Law (i.e., one-part). 
We conjecture that extra parameters in both of them lead to better results. 
\update{Yet, the use of these extra parameters is validated by significantly better AIC and BIC (positive to strong evidence seen from Table \ref{tab:AICBIC}).
More specifically, we suggest researchers to use the Two-part variation for distal pointing tasks, which shows significantly better performance (substantial evidence based on AIC values and strong to positive evidence based on BIC values) compared to the other models according to our results.}

\update{\subsection{Display Comparison}}

\subsubsection{User Performances: Movement Time and Error Rate}
Pointing on displays \update{which are less curved} are significantly faster than 2000R display \update{for both the angular and linear measures.} 
In addition, users can perform overall equivalent or even faster movements with Flat than curved displays.
For accuracy, pointing on less curved displays (6000R and Flat) leads to significantly lower error rates than the 2000R and 4000R displays \update{for the angular condition.
In contrast, 2000R and 4000R had significantly lower error rates than less curved displays (6000R and Flat) for linear condition.}
 \update{This suggests that less curved displays lead to faster selection times while the error rate for the selection largely depends on how the targets are placed (angular or linear measurement).}
\update{Our goal was to fill the gap of \emph{pointing performance comparison}.
For the linear condition 
This work hence reveals a trade-off between benefits (e.g., less region bias, facilitating visual search, etc.) and user performances.}

\update{\subsubsection{User Feedback}}
\update{Participants generally preferred curved displays over flat displays, especially for linear measures. 4000R and 6000R were preferred over 2000R for both linear and angular measures. 
Four participants found that the 2000R display required more head and body movements due to it being more curved, while others mentioned that the Flat display provided adequate target width for angular conditions.
Note that such feedback might not hold in a different context. 
We hypothesize that different results could arise when dealing with reading or identification tasks instead of an abstract target selection task - especially toward the edge of a large flat screen.}

\section{Limitations and Future work}

In this section, we highlight the limitations that we observed during the study and discuss potential future works.

The present work is motivated by the opportunity to explore curved displays without physical constraints (e.g., costs, etc.) and by leveraging VR capabilities.
However, VR has a main limitation, namely the limited field-of-view in head-mounted displays.
We used an HMD with a 90\degree field-of-view.
While our work provides indications of performance with virtual large curved displays, future studies could explore pointing performance between physical and virtual curved displays.
\update{Additionally, to improve generalization and diversity, we acknowledge the need for a broader range of participants in future studies.}
Furthermore, additional studies could investigate the effects of different field-of-views on HMDs on users' pointing performance.




In our user study, we use three Fitts law variations in order to explain movement times.
All three models demonstrate good predictions, even though they all intend to capture different aspects of pointing tasks. 
As our goal was to explore users' pointing performance with the standard factors in combination with relevant curve-related factors, we didn't investigate any formal information-theoretic models \cite{gori2018speed} for pointing tasks to better understand them.
Future work can adopt a more theoretical approach to explore curved display characteristics (e.g., curvature, position, or offset) and their relationship with the Fitts law variations. 
\update{Lastly, it could be interesting to study if users intuitively find the best locations to maximize the screen real-estate captured by their field-of-view on their own. 
Future work can explore scenarios less constrained than in this work to observe users' behavior and position in the VR environment.}


%% file: 11_conclusion.tex
\section{Conclusion}
In this paper, we investigated users pointing performance on large displays in VR. 
More specifically, we explored the effect of large display specificities (i.e., display curvature) on users' pointing performance.
Overall, higher display curvatures (such as 4000R, 6000R, and Flat) are faster and more preferable for distal pointing tasks compared to 2000R for both visual properties (linear widths and amplitudes) and their motor properties (angular widths and amplitudes).
\update{We found that the error rate on display with different curvature primarily depends on how the targets are placed on the display - whether linear or angular measurements are used.}
Moreover, we showed that based on only speed, flat displays offer better or comparable performance to curved displays on 1D pointing tasks. 
The advantages of the flat display subside when both speed and accuracy are incorporated  into the performance criteria (i.e., considering throughput).
According to our findings, we provided suggestions about how to leverage display specificities to improve users' performance on large virtual displays.